\documentclass[fleqn,usenatbib]{mnras}

\usepackage{newtxtext,newtxmath}

\usepackage{booktabs}
\usepackage{multirow}
\usepackage{orcidlink}

\DeclareRobustCommand{\VAN}[3]{#2}
\let\VANthebibliography\thebibliography
\def\thebibliography{\DeclareRobustCommand{\VAN}[3]{##3}\VANthebibliography}

\usepackage{graphicx}	
\usepackage{amsmath}

\title[An X-ray reflection study of H1821$+$643]{Evidence for a moderate spin from X-ray reflection of the high-mass supermassive black hole in the cluster-hosted quasar H1821$+$643}

\author[J. Sisk-Reynés et al.]{Júlia Sisk-Reynés \orcidlink{0000-0003-3814-6796},$^{1}$\thanks{Contact details: jms332@cam.ac.uk.}
Christopher S. Reynolds \orcidlink{0000-0002-1510-4860},$^{1}$
James H. Matthews \orcidlink{0000-0002-3493-7737},$^{1}$ \newauthor
and Robyn N. Smith \orcidlink{0000-0001-5626-5209}$^{2}$\\
$^{1}$ Institute of Astronomy, University of Cambridge, Madingley Road, Cambridge CB3 OHA, UK\\
$^{2}$ Dept. of Astronomy, University of Maryland, College Park, MD 20742, USA\\
}

\date{\today}

\pubyear{2022}

\begin{document}
\label{firstpage}
\pagerange{\pageref{firstpage}--\pageref{lastpage}}
\maketitle
\begin{abstract}
We present an analysis of deep \textit{Chandra} Low-Energy and High-Energy Transmission Grating archival observations of the extraordinarily luminous radio-quiet quasar H1821$+$643, hosted by a rich and massive cool-core cluster at redshift $z=0.3$. These datasets provide high-resolution spectra of the AGN at two epochs, free from contamination by the intracluster medium and from the effects of photon pile-up, providing a sensitive probe of the iron-$K$ band. At both epochs, the spectrum is well described by a power-law continuum plus X-ray reflection from both the inner accretion disc and cold, slowly-moving distant matter. Adopting this framework, we proceed to examine the properties of the inner disc and the black hole spin. Using Markov chain Monte Carlo (MCMC) methods, we combine constraints from the two epochs assuming that the black hole spin, inner disc inclination, and inner disc iron abundance are invariant. The black hole spin is found to be modest, with a 90\% credible range of ${a}^{*}=0.62^{+0.22}_{-0.37}$; and, with a mass $M_\mathrm{BH}$ in the range $\log (M_\mathrm{BH}/M_\odot)\sim 9.2-10.5$, this is the most massive black hole candidate for which a well-defined spin constraint has yet been obtained. The modest spin of this black hole supports previous suggestions that the most massive black holes may grow via incoherent or chaotic accretion and/or SMBH-SMBH mergers.
\end{abstract}

\begin{keywords}
galaxies: active -- galaxies: quasars: individual: H1821$+$643 --- galaxies: quasars: supermassive black holes -- X-ray: galaxies -- X-rays: individual: H1821$+$643
\end{keywords}

\section{Introduction}
\label{sec:s1_intro}

Galaxy clusters are the most massive bound structures in the Universe. In addition to forming galaxies, most of the baryons in clusters reside in the intracluster medium (ICM), a hot plasma ($\sim10^{8} \ \mathrm{K}$) that is in hydrostatic equilibrium with the dark matter potential well providing $\sim85\%$ of any cluster's mass budget. 

Most cool-core clusters host a central brightest cluster galaxy (BCG), which itself hosts an active galactic nucleus (AGN). Typically, the ICM cores of relaxed clusters represent the bulk of the X-ray emission and have cooling timescales $<1 \ \mathrm{Gyr}$, which, if left unchecked, should result in star formation rates that exceed those that have been observed in nearby clusters \citep[for a recent overview, see][]{McDonald_2018_coolingFlowProblem}. This disparity is one of the by-products of the cooling flow problem, that is, the unexpected lack of star formation within BCGs compared to the star formation rate one would predict from typical ICM cooling timescales.  

The most widely accepted explanation of the cooling flow problem is mechanical heating of the ICM by relativistic jets released by cluster-hosted AGNs. The presence or absence of these jets forms the basis of the radio-loud/radio-quiet AGN classification scheme. In fact, jets in radio-loud Active Galactic Nuclei interact with and thereby impact the thermal structure of the ICM \citep[e.g., see][]{jetsAsEvidenceForAGNdiscovery_fabian_2012, 2015_julie_selectedSPTclusters_jets_inflatingxrayBubbles}. In several relaxed clusters, the thermal properties of the ICM can be probed by studying filaments -- or optical-emission nebulae -- that are believed to contain information about mergers their host BCGs may have undergone. Nevertheless, the mystery behind the origin and formation history of BCGs remains unresolved and, in general, the AGN-ICM feedback cycle in radio-quiet AGNs is also poorly understood.

H1821+643 is an extraordinarily luminous radio-quiet quasar centred within a massive \citep[i.e. of mass $6.3 \times 10^{14}M_\odot$,][]{Planck13_tSZCatalogue} cool-core galaxy cluster, and has redshift $z = 0.299$ \citep[][]{ArticleComputing_z_0.299_ForH1821+643}. In 1991, \textit{Ginga} and \textit{IUE} observations of this quasar (at the time, unknown to be hosted by such a rich cluster) detected an iron-$K$ emission line centred at $(6.6 \pm 0.3) \ \mathrm{keV}$ in the quasar rest frame \citep[see][]{1991ApJ_ginga_ironEmissionLine_e1821+643, 1991_UVstudy}. Moreover, \textit{ROSAT PSPC} observations were the first to detect clear X-ray emission from the host cluster. A detailed study of the ICM and its thermal properties was presented in \citet[][]{Russell10}, who performed a \textit{Chandra} imaging study with a modest exposure ($85 \ \mathrm{ks}$) to spatially isolate the cluster emission from that of the powerful AGN. 

The quasar bolometric luminosity between $2-10 \ \mathrm{keV}$ (rest energies), estimated from its observed X-ray luminosity ($\sim 10^{45} \ \mathrm{erg/s}$), was computed to be $L_\mathrm{bol} = 2 \times 10^{47} \ \mathrm{erg/s}$ \citep[][]{Russell10}. The quasar does \textit{not} present signs of jet emission and was therefore categorised as radio-quiet \citep[][]{BLUNDELLandRawlings_e1821+643_2001}. Interestingly, this AGN was linked to an FR-I radio source extending to a maximum size of $280 \ \mathrm{kpc}$ beyond the central engine \citep[][]{BLUNDELLandRawlings_e1821+643_2001}. Moreover, the detection of diffuse radio emission within the cluster volume, with a largest linear size of $1.1 \ \mathrm{Mpc}$, was attributed to the existence of a radio halo co-centred with the cluster \citep[][]{CL1821+643_Bonafede14, Kale16_CL1821+643}. This is an exceptional finding for a cool-core cluster, since radio haloes are typically attributed to merging clusters. Although unclear, the radio halo within CL1821+643 is thought to have originated via cluster-cluster merger events prior to the cluster's relaxation \citep[][]{CL1821+643_Bonafede14}.

The thermal profiles of the host cluster, CL1821$+$643 -- inferred from a set of consecutive deprojected spectra of the intrinsic cluster emission -- confirmed its generic cool-core attributes \citep[][]{Russell10}. The ICM pressure and density distributions are centrally-peaked with pronounced rises within $\sim30 \ \mathrm{kpc}$ of the cluster centre. \citet[][]{Walker_h1821PressureProfile} noted that CL1821$+$643 possesses an anomalous low entropy core up to $80 \ \mathrm{kpc}$ when compared to other relaxed clusters of similar masses, suggesting that recent quasar activity has had a relevant role in affecting the ICM in its immediate vicinity. Interestingly, \citet[][]{Russell10} showed that, within $5 \ \mathrm{kpc}$ of the cluster/BCG centre, Compton cooling of the ICM by the quasar radiation prevails over the usual Bremsstrahlung cooling, leading to a Compton cooling powered cooling flow \citep[][]{1990_fabianANDcrawford_agn_compton_cooling_agn}. Moreover, the large-scale properties of the ICM were found to remain unaffected by quasar activity \citep[][]{Russell10}.

The SMBH centred within CL1821+643 is believed to be one of the most massive SMBHs in the local Universe, whose mass $M_\mathrm{BH}$ has been estimated through a variety of methods. To begin with, both \citet[][]{kim+08_spitzer_original} and \citet[][]{floyd_bcg_mass} used archival \textit{HST} observations and fitted the bulge luminosity of the central galaxy to derive an estimate of $M_\mathrm{BH} \sim (1, 3)\times {10}^{9}M_\odot$, respectively, based on an assumption between the bulge luminosity-SMBH mass. Additionally, two independent studies provided an estimate of $M_\mathrm{BH}\sim 3\times {10}^{9}M_\odot$ by using a thermal accretion disc model to explain the hump-like features present in the quasar’s spectral energy distribution in the UV band \citep[][]{Kolman_1991_UV_thinDiscModel,Shapovalova_Monitoring_binBHcandidate}. This result was consistent with a mass estimate computed on the basis of a single-epoch H$\beta$ emission line measurement \citep[][]{capellupo_1821+643}. Moreover, two independent X-ray studies inferred significantly higher mass values, as follows. Firstly, \citet[][]{Reynolds14} reported an upper mass estimate of $6\times {10}^9 M_\odot$ from a $375 \ \mathrm{ks}$ \textit{Suzaku} X-ray reflection study based on the relation between the ionisation state of the accretion disc and the Eddington fraction \citep[][]{Reynolds14}. Secondly, \citet[][]{Walker_h1821PressureProfile} suggested a SMBH mass of $3\times {10}^{10}M_\odot$, based on the assumption that this system had been locked into a Compton-cooled feeding cycle. If indeed possible, these cycles could result in the growth of very massive SMBHs with $\mathrm{log}(M_\mathrm{BH}/M_\odot)\gtrsim 10$, hence having important implications for our understanding of the AGN-ICM feedback cycle in such Active Galactic Nuclei.

A programme of \textit{Chandra} observations of H1821$+$643 were taken in 2001 with the Low-Energy and High-Energy Transmission Gratings (LETG and HETG, respectively), principally to search for quasar absorption lines from the Warm-Hot Intergalactic Medium \citep[WHIM, see][]{Fang_2002_WHIMObs_h1821+643, Mathur_2003_WHIM_h1821+643, Yaqoob_2005, Kovacs_LETGChandraObs_MissingBaryons}. In addition to providing a high-resolution spectrum of the intrinsic AGN emission with minimal cluster contamination, the use of \textit{Chandra}'s grating instruments provided a successful route to circumventing photon pile-up. Recently, we combined these same LETG/HETG spectra to set strong bounds on the coupling of very light Axion-Light Particles (ALPs) to electromagnetism, setting an upper bound on the ALP-photon constant of $g_\mathrm{a\gamma} < 6.3 \times {10}^{-13} \ {\mathrm{GeV}}^{-1}$ for ALP masses $\lesssim {10}^{-12} \ \mathrm{eV}$ at 99.7\% confidence \citep[][]{sisk21_alps}. The previous analysis of these data most pertinent to our current study was presented by \citet[][]{Yaqoob_2005}. These authors used the LETG/HETG data to study the iron-K emission line, finding it to be
broadened and identifying a possible absorption line. Importantly, the \citet[][]{Yaqoob_2005} study highlighted
the ability of the LETG to study the iron-K band.

H1821+643 is one of the most massive SMBHs whose angular momentum $J$ has been constrained, albeit weakly until now \citep[see Fig. 6 of][]{reynolds2019ObsBlackHoleSpinsReview}. Historically, $J$ has been quantified through a dimensionless spin parameter ${a}^{*}$ defined as ${a}^{*} = cJ/{G M_\mathrm{BH}^{2}}$, where $M_\mathrm{BH}$ is the black hole mass, $c$ is the speed of light in the vacuum and $G$ is Newton’s gravitational constant. Importantly, ${a}^{*}$ can range between $-0.998$ to  $+0.998$, with ${a}^{*}<0$ and ${a}^{*}>0$ indicating that the black hole spins in either a retrograde or prograde sense to the matter accreted onto it, respectively.

At a fundamental level, the no-hair theorem of General Relativity dictates that the mass and the spin completely determine the spacetime structure around astrophysical (uncharged) black holes. Therefore, analysing observational bounds on ${a}^{*}$ across the widest possible range of black hole masses is clearly interesting for fundamental physics. Additionally, spin measurements provide a window for constraining black hole formation and growth models. Specifically, most observational studies of low-mass SMBHs, i.e. of $\mathrm{log}(M_\mathrm{BH}/M_\odot)\sim 6 - 7$, have found most of these sources to have maximal or extreme spins (${a}^{*}\gtrsim 0.9$), suggesting that they grow following coherent accretion events, as well as episodes of perfect isotropic gas fuelling into the AGN \citep[][]{Volonteri_2005_smbh_massiveBHs_growth}. In coherent accretion, the angular momentum of the accreted material aligns with and therefore contributes to increasing ${a}^{*}$. However, above masses of $\mathrm{log}(M_\mathrm{BH}/M_\odot)\sim 7$, a population of black holes with more moderate spin parameters (centred around ${a}^{*}\sim 0.5-0.7$) appear (see Fig. \ref{figure:updated_spin_vs_mass_plot_reynolds20}). This is in line with both semi-analytic and hydrodynamical cosmological models which suggest that, for the most massive black holes, incoherent accretion and/or SMBH-SMBH merger scenarios are more relevant growth channels \citep[see Sec 5.1 of][and references therein]{reynolds2019ObsBlackHoleSpinsReview}. Incoherent or chaotic accretion events arise when the spin orbits of the accreted material and that of the SMBH are antialigned, which lowers the magnitude of ${a}^{*}$.

H1821$+$643 was tentatively constrained to have spin ${a}^{*} > 0.4$ by \citet[][]{Reynolds14} on the basis of using the \texttt{relxill\_lp} relativistic X-ray reflection code \citep[][]{2014_dauserANDgarcia_RoleofReflFrac, GarciaANDdauser_2014_relxill_rayTracing} to describe the soft excess resulting from a deep \textit{Suzaku} observation. \texttt{relxill\_lp} describes the relativistically blurred reflection spectrum from a geometrically thin and optically thick accretion disc around a Kerr black hole that is steadily illuminated by a hot ($\sim 10^9 \ \mathrm{K}$) compact plasma (the X-ray corona) located along its spin axis. The most obvious signature of such a reflection component is the appearance of the Fe-$K\alpha$ line, with a rest frame energy of $6.4 - 6.97 \ \mathrm{keV}$ depending on the ionisation state of the disc. We refer the reader to  \citet[][]{1991_haart_maraschi_Seyfert_twoPhaseDiscs,1993_haart_maraschi_XraySpectra_twoPhase, Fabian1989_Diskline} and \citet[][]{Ballanytyne_2001_XrayRefl_innerAccDiscs} for further details on the nature of the X-ray corona and of the broadened iron line in AGN. A rigorous analysis of the iron-$K$ band in AGN provides bounds on several fundamental parameters of the system including the black hole spin ${a}^{*}$ and iron abundance $A_\mathrm{Fe}[Z_\odot]$ of the disc. For a wide range of ionisation states, the reflection spectrum from the inner disc also contains a forest of soft X-ray emission which can be broadened into a pseudo-continuum soft excess. Due to ICM contamination and instrumental calibration issues, \citet[][]{Reynolds14} did not isolate the broad iron line in H1821+643, instead deriving their lower spin bound by using \texttt{relxill\_lp} for a phenomenological description of the soft excess.

In addition, \citet[][]{Reynolds14} detected a $6.4 \ \mathrm{keV}$ emission line attributed to circumnuclear material in the vicinity of the quasar, characterised to have a sub-solar iron abundance ($\sim 0.4Z_\odot$). Yet again, this represents an unexpected finding when comparing H1821$+$643 to similar-mass SMBHs whose surrounding cold gas present solar and mildly super-solar metallicities \citep[refer to the findings of several Seyfert-1 \textit{Suzaku} spectra in][]{patrick_suzaku_spectra_of_seyfert1agn}. 

In this work, we reanalyse the archival LETG/HETG \textit{Chandra} observations of H1821$+$643 to constrain its physical properties. With an observed energy range of $1.5 - 8.5 \ \mathrm{keV}$ (rest frame $2 - 11 \ \mathrm{keV}$), we find statistical evidence for relativistic X-ray reflection of the primary source (or corona) from the inner accretion disc. Crucially, we clearly identify and can model the broadened iron line in these spectra. Our combined constraints are found by merging the output of Markov chain Monte Carlo (MCMC) chains performed when fitting the LETG/HETG spectra separately. At $90\%$ confidence, we predict the inclination $i$ and metallicity of the accretion disc to be $i\sim {41}^{\circ} - {48}^{\circ}$ and $A_\mathrm{Fe}\sim (0.6 - 2.4)Z_\odot$.

This paper is organised as follows. In Section \ref{sec:s2_obs_and_dataRed}, we present the archival spectroscopic \textit{Chandra} observations of H1821+643 we have employed. In Section \ref{sec:s3_fitting_spectral_withPlots}, we use relativistic reflection models to describe the individual LETG/HETG spectra. We then present the best-fit model parameters inferred when describing the separate LETG/HETG spectra with the \texttt{relxill\_lp} relativistic reflection model, in addition to considering the Fe-$6.4 \ \mathrm{keV}$ emission line from circumnuclear material in the immediate vicinity of the AGN, as well as the effects of Galactic absorption. In Sec. \ref{sec:s4_Results_Physical_Params}, we introduce the MCMC chains performed on the individual HETG/LETG spectra and combine them following the Bayesian framework presented in Sec. \ref{sec:s5_statisticalFramework_jointConstranits}. The convergence of all MCMC chains presented in this document was validated with the Geweke diagnostic \citep[][]{Geweke_chainDiagnostics}. We discuss our results and conclusions in Secs. \ref{sec:s6_discussion} and \ref{sec:s7_conclusions}, respectively. At $90\%$ confidence, we constrain the spin ${a}^{*}$, and inclination $i$ and iron abundance $A_\mathrm{Fe}$ of the accretion disc for this remarkable system. Our study excludes retrograde, non-rotating and maximal spins, supporting black hole growth scenarios that favour late incoherent or chaotic accretion events for black hole masses $\gtrsim 10^8M_\odot$ \citep[][]{king_pringle_chaotic, Zhang2019_and_Lu_spin}, and/or SMBH-SMBH merging events \citep[][]{bustamante_springel_illustrisTNG} up to masses $\gtrsim {10}^{10}M_\odot$.

Throughout this paper, we assume a flat ($\Omega_\mathrm{\kappa} = 0.0$), $\Lambda$CDM cosmology: $H_0 = 70 \ \mathrm{km/s/Mpc}$, $\Omega_\mathrm{m} = 0.3$, $\Omega_\mathrm{\Lambda} = 0.7$ \citep[][]{PlanckCollaborationLatestResults18}. With a redshift of $z=0.299$ \citep[][]{ArticleComputing_z_0.299_ForH1821+643}, this translates into a quasar luminosity distance of $1.55 \ \mathrm{Gpc}$. We assume a neutral hydrogen column density local to the Milky Way of $N_\mathrm{H} = 3.51 \times 10^{20} \ \mathrm{cm^{-2}}$ \citep[][]{Reference_forNH}, using the element abundance ratios of \citet[][]{1989_XspecAng_abundances}. We perform the most rigorous statistical analysis on the X-ray reflection spectrum of this source to date by using MCMC tools provided in the \textsc{xspec} X-ray Spectral Fitting Package \citep[v12.11.1,][]{XSPEC_1996Arnaud}. Based on previous literature, we assume that the separate AGN spectra do \textit{not} present signatures characteristic of warm or partially-covering absorbers, nor those attributed to the presence of AGN winds released into the ISM in the form of ionised outflows \citep[][]{Fang_2002_WHIMObs_h1821+643, Mathur_2003_WHIM_h1821+643,Oegerle_2000_FUSEObs_confirm_no_outflowEvidence}.

\section{Observations and data reduction}
\label{sec:s2_obs_and_dataRed} 

The ObsIDs of the Low-Energy Transmission Grating (LETG) observations of H1821$+$643 we have employed are outlined in Tab. \ref{table:Tab1_obsIds_reduced_clean_Chandra_data}. The LETG spectra were read out by the Advanced CCD Imaging Spectrometer-S array (ACIS-S). These observations, taken between 17-Jan-2001 and 24-Jan-2001, add up to a cleaned time exposure of $471.4 \ \mathrm{ks}$. Separately, we also use the only High-Energy Transmission Grating (HETG) observation of the source (refer to Tab. \ref{table:Tab1_obsIds_reduced_clean_Chandra_data}). This grating observation, also read on ACIS-S, began on 09-Feb-2001. Note that the HETG consists of the High-Energy and Medium-Energy gratings (HEG and MEG, respectively), providing two simultaneous views of the AGN spectrum. 
\begin{table}
\centering 
\renewcommand{\arraystretch}{1.2}
\begin{tabular}{c c c}
\toprule 
\midrule
  Grating & ObsID & Exposure \\ 
  \hline \hline
  LETG & 2186 & 165.4 $\mathrm{ks}$ \\ 
\cmidrule(l){2-3} %{2-2}
  & 2310 & 163.7 $\mathrm{ks}$ \\
\cmidrule(l){2-3}
  & 2311 & 90.5 $\mathrm{ks}$ \\
\cmidrule(lr){2-3}
  & 2418 & 51.8 $\mathrm{ks}$ \\
  \hline
  HETG & 1599 & 99.6 $\mathrm{ks}$ \\ \hline \hline
\end{tabular}
\caption{\label{table:Tab1_obsIds_reduced_clean_Chandra_data}ObsIDs and cleaned exposure times of the achival set of LETG/HETG \textit{Chandra} observations of H1821$+$643 we have employed.} 
\end{table}

Armed with these high-quality grating spectra, combined with the order-sorting performed by ACIS-S, the AGN emission can be isolated from that of the cluster. The archival \textit{Chandra} observations were reduced and reprocessed with CIAOv4.13 and CALDBv4.9.4, following the data reduction procedures described in Sec. 2.1 of \citet[][]{sisk21_alps}. The background spectral file generated with \textsc{CALDB}'s \texttt{bkgpha} command was Poisson-distributed. This background was treated with a Poisson profile likelihood (\texttt{wstat}) throughout the entirety of our spectral analysis.

As presented in Sec. \ref{sec:s3_fitting_spectral_withPlots}, we fit the LETG in the $1.5 - 6.5 \ \mathrm{keV}$ band, the MEG in the $1.5 - 7.0 \ \mathrm{keV}$ band, and the HEG in the $1.5 - 8.5 \ \mathrm{keV}$ band (observer frame). The choice of a 1.5\,keV lower bound is driven by the desire to avoid additional soft emission that may be associated with a very centrally-peaked Compton-cooled ICM core. The upper bounds of each grating array are determined by the high-energy response and limited photon statistics. The resulting X-ray luminosity and flux for each of the two epochs of data are listed in Tab. \ref{table:bandLumi_and_fluxes}. 
\begin{table*}
\centering 
\renewcommand{\arraystretch}{1.2}
\begin{tabular}{c c c}
\toprule 
\midrule
 Magnitude & LETG & HETG \\ \hline \hline
 Fitted energies (rest frame) & $2.0 - 8.5 \ \mathrm{keV}$ & $2.0 - 9.0 \ \mathrm{keV}$ (MEG), $2.0 - 11.0 \ \mathrm{keV}$ (HEG) \\ \hline
 Cleaned exposure time & $471.4 \ \mathrm{ks}$ & $99.6 \ \mathrm{ks}$ \\ \hline
 $L_\mathrm{2 - 10 \ \mathrm{keV, \ rest}}$ & $3.23\times 10^{45} \ \mathrm{erg}/ \mathrm{s}$ & $3.54\times 10^{45} \ \mathrm{erg}/ \mathrm{s}$ \\ \hline
 $F_\mathrm{2 - 10 \ \mathrm{keV, \ observed}}$ & $1.16 \times 10^{-11} \ \mathrm{erg/{cm}^{2}/ {s}}$ & $1.28\times 10^{-11} \ \mathrm{erg/{cm}^{2}/ {s}}$ \\ \hline\hline
\end{tabular}
\caption{\label{table:bandLumi_and_fluxes}Outline of the rest energies fitted in the individual LETG/HETG spectra analysed, as well as their corresponding total clean exposures. We include the unabsorbed band luminosity ($L$, rest frame) and flux ($F$, observer frame) within $2 - 10 \ \mathrm{keV}$, where the values quoted for the HETG correspond to the averaged quantities between the HEG/MEG.}
\end{table*}

Given their different times of observation and thereby the possibility of intrinsic AGN variability, we perform independent analyses of the LETG and HETG spectra. In both cases, we recover a high-resolution AGN spectrum free from pile-up with which we can perform high quality modelling. 

\section{Relativistic reflection model fitting}
\label{sec:s3_fitting_spectral_withPlots}

In Secs. \ref{sec:s3.1_fitting_letg} and \ref{sec:s3.2_fitting_hetg}, we show how the LETG/HETG datasets present robust evidence for spectral features above and beyond the typical power-law continuum expected from direct coronal emission, as well as Galactic absorption along the quasar line-of-sight. Our spectral modelling then proceeds under the assumption that these features come from the reflection of coronal emission from: I. An ionised inner accretion disc; and II. Slowly-moving neutral matter surrounding the immediate vicinity of the AGN. We restricted our spectral analysis to photon energies $\geq 2 \ \mathrm{keV}$ (rest energies), given the unknown nature of the soft excess in this source \citep[refer to][]{Reynolds14}.

In all spectral models mentioned below, we include the effects of Galactic absorption using the \texttt{tbabs} Tuebingen-Boulder model \citep[][]{TbAbs_References}. We assume a constant Galactic neutral hydrogen column density of $N_\mathrm{H} = 3.51 \times 10^{20} \ \mathrm{cm}^{-2}$ for all datasets \citep[][]{Reference_forNH}. 

\subsection{Fitting the LETG data}
\label{sec:s3.1_fitting_letg}

\begin{figure}
 \center
 \includegraphics[width=0.5\textwidth]{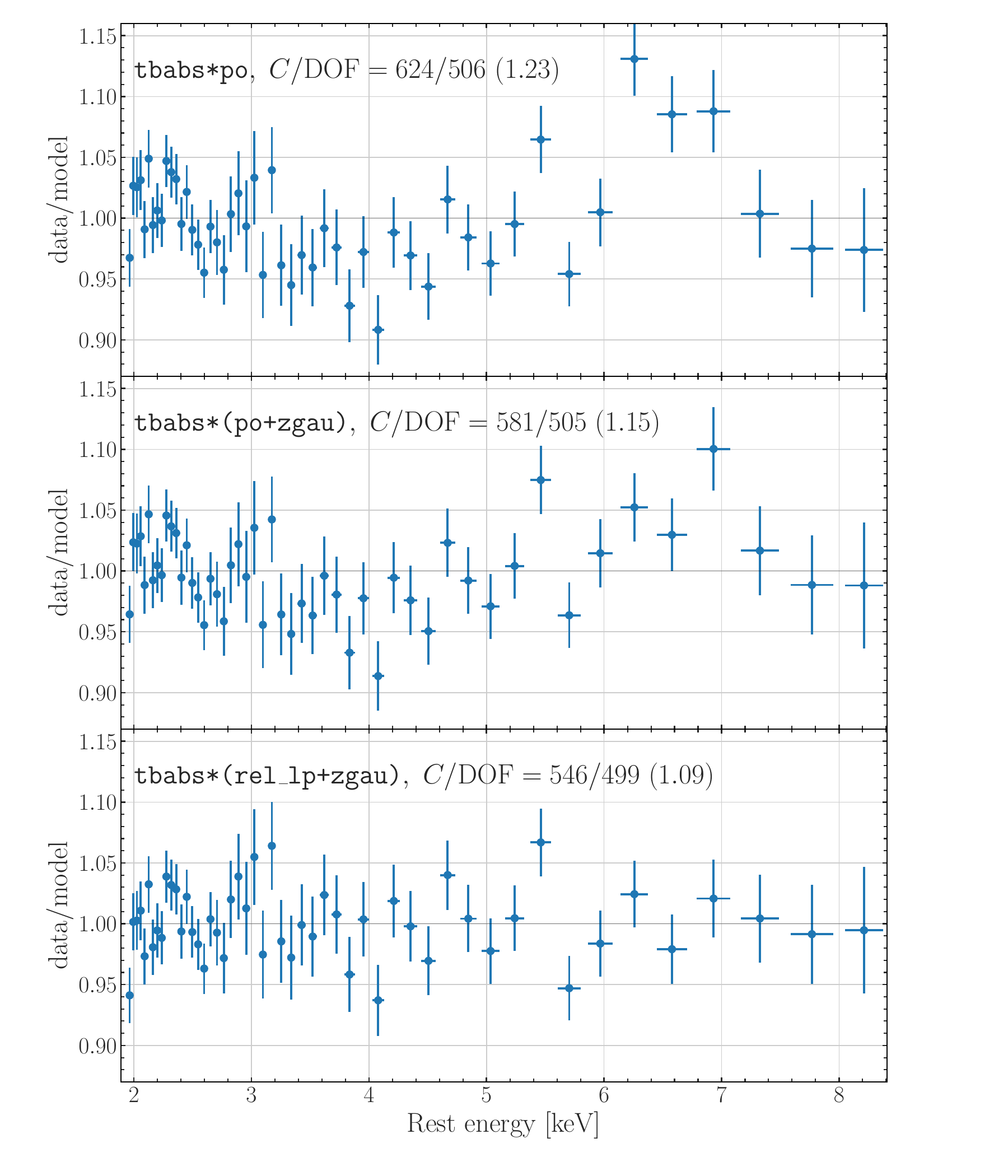}
 \caption{Ratio of the LETG data to the best-fit models outlined in Tab. \ref{tab:bestfits_relxilllp__from_cstat_min} (where \texttt{rel\_lp} denotes \texttt{relxill\_lp}). All error bars represent uncertainties within $1\sigma$. The $C$-stat and reduced $C$-stat values of a given model are displayed in each panel accordingly, and all models include a Galactic neutral hydrogen column density of $N_\mathrm{H} = 3.51 \times {10}^{20} \ \mathrm{cm}^{-2}$. For plotting purposes, all spectral data were re-binned to a target signal-to-noise ratio of 200 with the restriction that no more than 10 spectral bins were co-added.}
 \label{figure:residuals_xspec_letg_bestFit}
\end{figure}

Our spectral study is based on the minimisation of the \textit{Cash}-statistic (or \textit{C}-stat), a goodness-of-fit parameter quantifying how suitable any model is for describing Poisson-like distributed data \citep[][]{kaastra_cstat}. To begin with, we fitted the LETG spectrum with a single power-law (\texttt{po}). This power-law continuum, ubiquitous in AGN spectra, is attributed to the inverse Compton scattering of thermal disc photons by hot (i.e. of temperatures $\sim 10^{9} \ \mathrm{K}$) electrons in the disc corona. At $90\%$ confidence level ($90\% \ \mathrm{CL}$), this yields a spectral index for the primary continuum of $\textit{$\Gamma$} = 1.93\pm0.02$, typical of radio-quiet quasars \citep[][]{asca_spectra}, and a normalisation of $(4.0\pm0.1) \times 10^{-3} \ \mathrm{counts/s/keV}$ at 1\,keV. However, this fit leaves obvious unmodelled structure (see upper panel of Fig. \ref{figure:residuals_xspec_letg_bestFit}), most notably a narrow line at $6.4 \ \mathrm{keV}$ and broad residuals in the $5.5 - 8.4 \ \mathrm{keV}$ band (both quoted in the rest frame).

The $6.4 \ \mathrm{keV}$ line is readily identified as the iron-$K\alpha$ fluorescence line and has already been noted in this spectrum by \citet[][]{Fang_2002_WHIMObs_h1821+643}, as well as being detected by \textit{Newton-XMM} \citep[][]{jimenezBailon:2006_xmmNewton_6.4keV} and \textit{Suzaku} \citep[][]{Reynolds14}. This arises from irradiated, low-velocity circumnuclear (i.e. cold) material around the quasar. Adding a narrow Gaussian centred at $6.4 \ \mathrm{keV}$ to the single \texttt{po} model improved the LETG fit by $\Delta C = -43$ (see Tab. \ref{tab:bestfits_relxilllp__from_cstat_min}). However, unmodelled structure at energies within the $5.5 - 8.4 \ \mathrm{keV}$ band still remains (refer to central panel of Fig. \ref{figure:residuals_xspec_letg_bestFit}). Guided by the hypothesis that this may correspond to a relativistically broadened iron line from the ionised inner disc, we proceeded to describe the spectrum using the \texttt{relxill\_lp} reflection model. 

The \texttt{relxill\_lp} model describes the strength and shape of the iron line via the following free parameters: the height of the primary source $h[R_\mathrm{g}]$ (or corona), assumed to have a lamppost geometry, and the reflection fraction $\mathcal{R}$; the black hole spin, in dimensionless units, ${a}^{*}$; and the inclination $i[^{\circ}]$, the logarithm of the ionisation parameter $\mathrm{log}(\xi[\mathrm{erg~cm~s^{-1}}])$ and iron abundance $A_\mathrm{Fe}[Z_\odot]$ of the inner disc \citep[][]{2014_dauserANDgarcia_RoleofReflFrac, GarciaANDdauser_2014_relxill_rayTracing}. The ionisation parameter, $\xi = 4 \pi F_\mathrm{in} / n_\mathrm{disc}$, is a measure of the total incident flux on the disc $F_\mathrm{in}$ weighted by the disc density $n_\mathrm{disc}$. In \texttt{relxill\_lp}, $n_\mathrm{disc}$ is fixed to $10^{15} \ {\mathrm{cm}}^{-3}$. $\mathcal{R}$ is defined as the ratio of light illuminating the inner disc to that escaping to infinity and is quantified in the source rest frame. The flux irradiating the inner disc is inversely proportional to $h[R_\mathrm{g}]$, corresponds to direct coronal emission and is quantified through the usual power-law index $\Gamma$.

We use \texttt{relxill\_lp} to describe the disc reflection spectrum of H1821$+$643, assuming that the accretion disc extends down to the innermost stable circular orbit, $R_\mathrm{ISCO}$, and extends out to $400R_\mathrm{g}$ (where $R_\mathrm{g}$ is the gravitational radius of the black hole). In fact, provided that the height/extent of the corona is much less that the outer radius, the resulting reflection spectrum is insensitive to the choice of outer radius.

The position of the corona above the black hole is a free parameter in \texttt{relxill\_lp}. The primary continuum is modelled by a power-law cutoff at $300 \ \mathrm{keV}$ in the observed frame. The reflected component from the inner accretion disc is computed by splitting the inner disc into consecutive annuli at different radii to the black hole. The resulting incident (and thereby, reflected) coronal flux onto each annulus is therefore inversely proportional to the photon travel path associated with it. While \texttt{relxill\_lp} allows us to self-consistently tie the system geometry and reflection fraction, we chose not to impose this constraint (preferring to see concordance arise naturally). Thus we enabled the reflection fraction to be fitted freely. The resulting fit (reported in Table \ref{tab:bestfits_relxilllp__from_cstat_min}) is successful at describing the curvature present in the LETG, flattening the residuals $>5.5 \ \mathrm{keV}$ and further improving on the previous best-fit by a factor of $\Delta C = -35$ (see lower panel of Fig. \ref{figure:residuals_xspec_letg_bestFit}). Although the consideration of a lamppost geometry may seem oversimplistic, it serves to exclude non-physical regions of parameter space (e.g. reflection from an infinitely thin ring at the ISCO), as well as to set bounds on fundamental parameters of the system as follows. 

When describing the LETG with \texttt{relxill\_lp}, the spectral index of the primary continuum softens (with $\textit{$\Gamma$}=2.10\pm0.06$), compared to that found by the simple \texttt{po} model. This is unsurprising given that a steeper primary power-law component will naturally allow reflection for a good description of the spectrum at higher energies.
The \texttt{relxill\_lp} fit to the LETG data finds a spin of ${a}^{*}<0.79$ (90\% $\mathrm{CL}$), as well as a moderate reflection fraction, $\mathcal \sim 1.2 - 2.5$ \citep[consistent with the reflection fraction found by][]{Reynolds14}. In fact, this $90\% \ \mathrm{CL}$ for $\mathcal{R}$ agrees with the expectation from a lamppost geometry, given that maximal spins are excluded \citep[see Figs. 2-4 of][]{2014_dauserANDgarcia_RoleofReflFrac}, although we note that the height $h[R_\mathrm{g}]$ of the lamppost is poorly constrained. \subsection{Fitting the HETG data}
\label{sec:s3.2_fitting_hetg}
\begin{figure}
 \centering
 \includegraphics[width=0.5\textwidth]{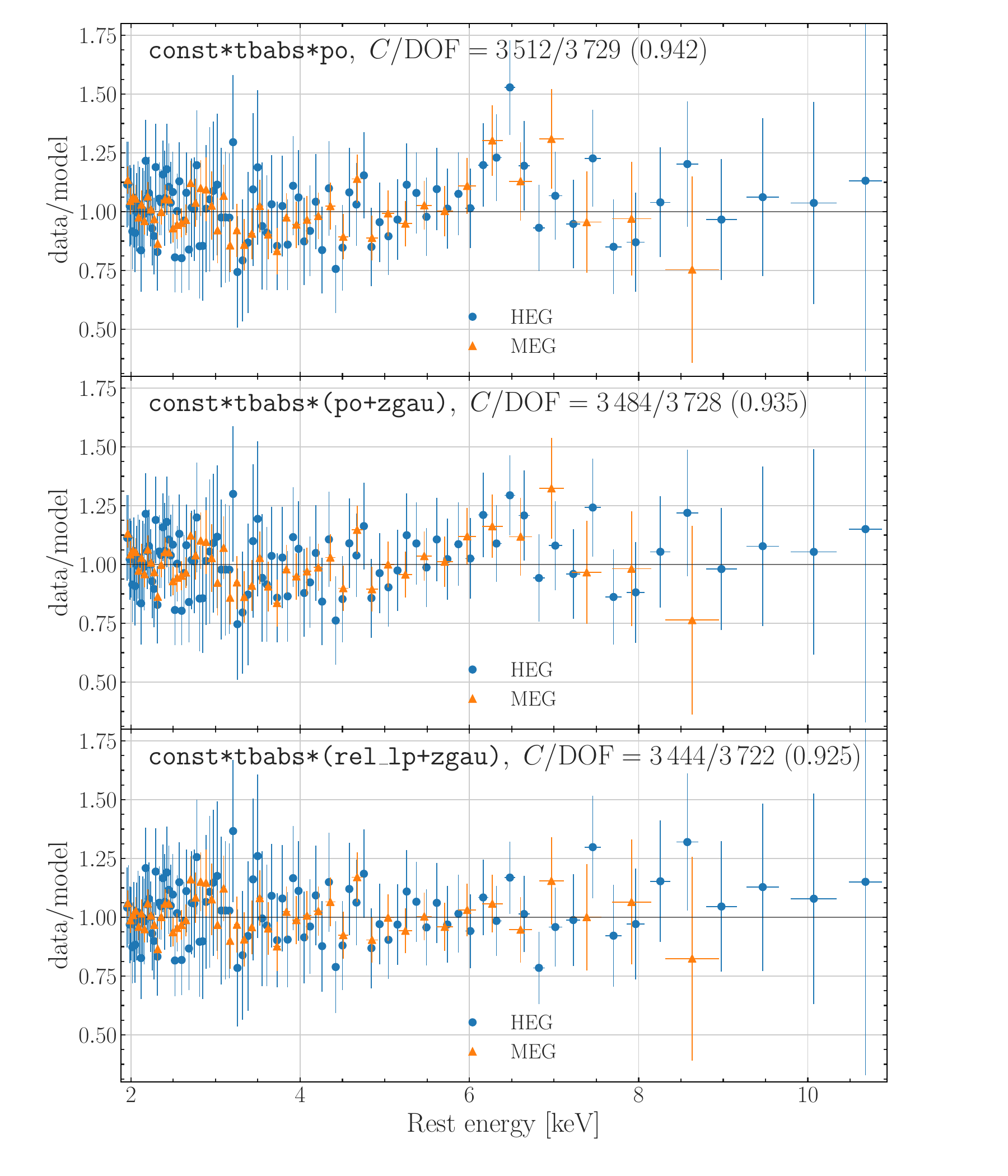}
\caption{Ratio of the HEG (blue) and MEG (orange) data to the best-fit models outlined in Tab. \ref{tab:bestfits_relxilllp__from_cstat_min} (where \texttt{rel\_lp} denotes \texttt{relxill\_lp}). All error bars represent uncertainties within $1\sigma$. The $C$-stat and reduced $C$-stat values of a given model are displayed in each panel accordingly, and all models include a Galactic neutral hydrogen column density of $N_\mathrm{H} = 3.51 \times {10}^{20} \ \mathrm{cm}^{-2}$. For plotting purposes, all spectral data were re-binned to a target signal-to-noise ratio of 200 with the restriction that no more than 25 spectral bins were co-added.}
 \label{figure:hetg_residuals_bestFit_CstatReduction}
\end{figure}

Our analysis of the HETG data closely follows that of the LETG described above, with the addition of accounting for potential flux-calibration uncertainties between the two grating arrays that constitute the HETG (that is, the HEG and MEG) via the inclusion of a multiplicative \texttt{const} component. When solely modelling the primary continuum with a power-law fit, we find a best-fit spectral index of $\textit{$\Gamma$} = 1.80 \pm 0.03$. This is noticeably harder than the best-fit $\textit{$\Gamma$}$ values provided by the LETG fits (see Tab. \ref{tab:bestfits_relxilllp__from_cstat_min}), suggesting that the source may have undergone intrinsic spectral change during the 16 days between the LETG and HETG observations. This gave a best-fit value of the HEG/MEG cross-calibration constant $1.04 \pm 0.03$. Out of all unmodelled residuals resulting from a simple \texttt{po} fit, the most distinct is the iron line at $6.4 \ \mathrm{keV}$ (see Fig. \ref{figure:hetg_residuals_bestFit_CstatReduction}). Again, we attributed this to reflection from a cold torus surrounding the outer accretion disc. The inclusion of a narrow Gaussian line at 6.4\,keV improves the goodness-of-fit by $\Delta C=-28$.

\begin{figure*}
 \center
 \includegraphics[width=0.75\textwidth]{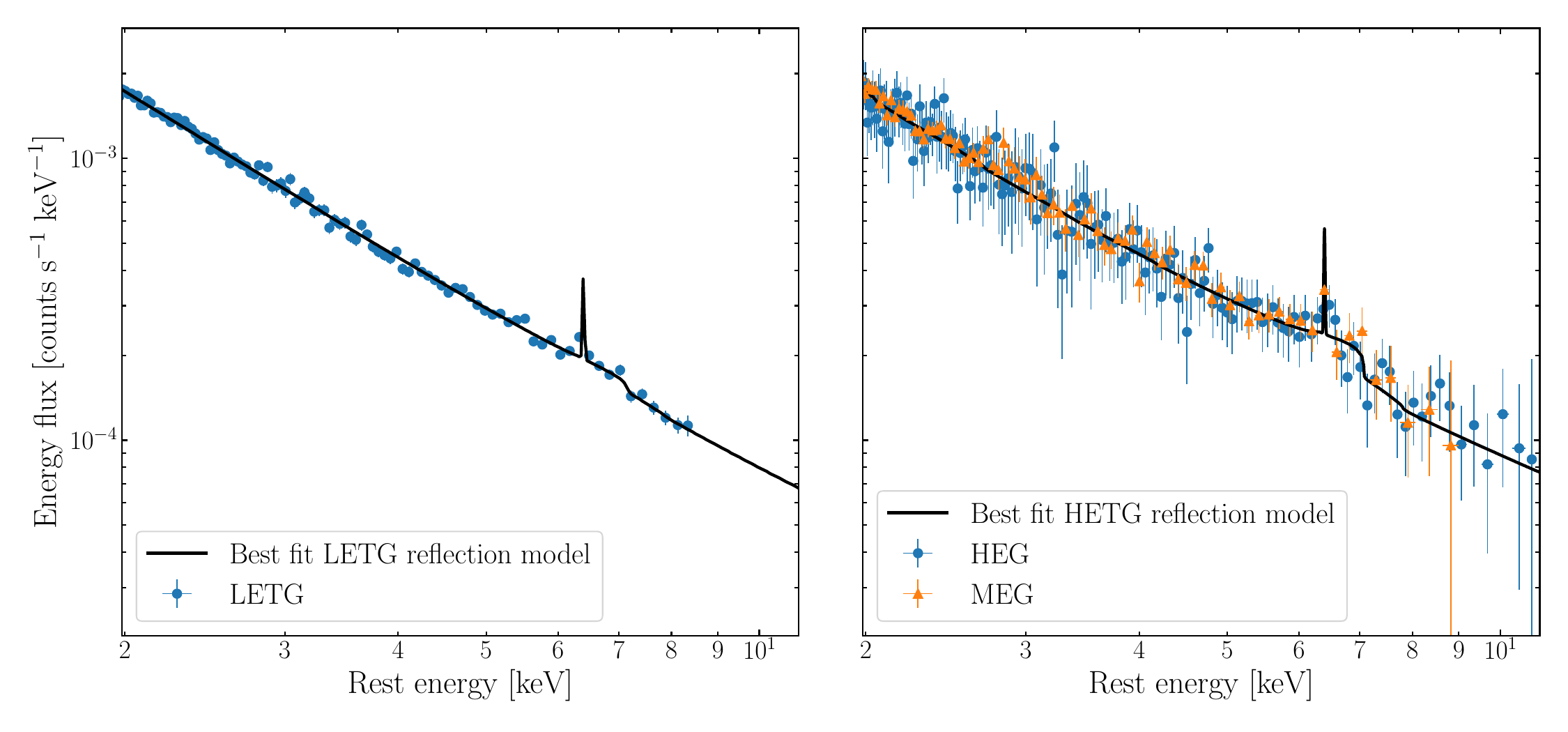}
 \caption{Effective area-corrected LETG (left) and HETG (right) spectral data employed throughout our analysis. For plotting purposes, all data were re-binned to a target signal-to-noise ratio of 100 with the restriction that no more than 5 and 15 spectral bins were co-added, respectively. The black lines show the best-fit relativistic+narrow reflection models outlined in Tab. \ref{tab:bestfits_relxilllp__from_cstat_min}. The effective area-corrected best-fit reflection model shown for the HETG dataset corresponds to that attributed to the HEG, since the best-fit model found for the MEG is equivalent to that of the HEG, but with its normalisation set by the best-fit cross-calibration constant, $1.04 \pm 0.02$.}
 \label{figure:bestFitModel_data}
\end{figure*}

As outlined by Tab. \ref{tab:bestfits_relxilllp__from_cstat_min}, we find statistical significance for relativistic X-ray reflection from the inner accretion disc in the HETG data when using the \texttt{relxill\_lp} model. Indeed, the latter improved the fit by $\Delta C = -40$ when compared to the \texttt{const*tbabs*(po+zgau)} model, with a best-fit value of the cross-calibration constant of  $1.04 \pm 0.02$. This model is shown by Fig. \ref{figure:bestFitModel_data}, which also illustrates the \texttt{tbabs*(relxill\_lp+zgau)} best-fitting regime found for the LETG dataset. As illustrated in Fig. \ref{figure:hetg_residuals_bestFit_CstatReduction}, \texttt{relxill\_lp} captures the underlying residuals present in the HETG data within the physically relevant iron band. Importantly, at $90\% \ \mathrm{CL}$, the HETG data finds a spin parameter in the range ${a}^{*}\sim 0.05-0.94$, allowing us to reject retrograde, non-rotating, and extreme (${a}^{*}\gtrsim 0.94$) spins.Additionally, the best-fit inclination values of the accretion disc $i$ inferred when describing the individual LETG and HETG datasets with \texttt{relxill\_lp} are reassuringly consistent. Despite being more broadly distributed in the HETG fit, this also applies to the iron abundance $A_\mathrm{Fe}$. 
\begin{table*}
\begin{center}
\renewcommand{\arraystretch}{1.5}
\begin{tabular}{lcccc}
\hline\hline
Spectral Model & LETG parameters & HETG parameters & LETG $C / \mathrm{DOFs} \ (C_\nu)$ & HETG $C / \mathrm{DOFs} \ (C_\nu)$ \\\hline

\texttt{po} & $\textit{$\Gamma$} = 1.93 ^{+0.02}_{-0.02}$ & $\textit{$\Gamma$} = 1.80^{+0.03}_{-0.03}$ & 624 / 506 (1.23) & 3\,512 / 3\,729 (0.942) \\\hline

\texttt{po} + & $\textit{$\Gamma$} = 1.93 ^{+0.02}_{-0.02}$ & $\textit{$\Gamma$} = 1.81^{+0.03}_{-0.03}$ & 581 / 505 (1.15) & 3\,484 / 3\,728 (0.935) \\
\texttt{zgau} & $\mathrm{EW} = 0.02 \ \mathrm{keV}$ & $\mathrm{EW} = 0.04 \ \mathrm{keV}$ & \\ \hline
\texttt{relxill\_lp} + & $\textit{$\Gamma$}=2.10^{+0.06}_{-0.06}(*)$ & $\textit{$\Gamma$} \sim 1.81 - 1.96$ & 546 / 499 (1.09) & 3\,444 / 3\,722 (0.925) \\
 & $ {h}^{\dagger} < 12R_\mathrm{g}$, $\mathcal{R} = 1.78^{+0.68}_{-0.61}(*)$ & $ h<7R_\mathrm{g}$, $\mathcal{R} = 3.94^{+3.71}_{-2.44}$ & \\
 & $i \sim {41^{\circ}} - {48^{\circ}}$ & $i \sim {38^{\circ}} - {56^{\circ}}$ & \\ 
 & ${a}^{*} < 0.79$ & ${{a}^{*}}^{\dagger} \sim 0.05 - 0.94$ & \\
 & $\mathrm{log}(\xi) \lesssim 1.26$ & $\mathrm{log}(\xi) \lesssim 3.07$ & \\ 
 & $A_\mathrm{Fe} \lesssim 1.68$ & $A_\mathrm{Fe}^{\dagger} \lesssim 4.78$ & \\
 \texttt{zgau} & $\mathrm{EW} = {26}^{+534}_{-24}\ \mathrm{eV}$ & $\mathrm{EW} = {30}^{+770}_{-29}\ \mathrm{eV}$ & \\ \hline \hline
\end{tabular}
\end{center}
\caption{\label{tab:bestfits_relxilllp_from_cstat_min}Best-fit parameters for the \texttt{po}, \texttt{po+zgau} and \texttt{relxill\_lp+zgau} models when fitted, separately, to the LETG and HETG \textit{Chandra} data. All error ranges are quoted at the $90\%$ confidence level ($90\% \ \mathrm{CL}$). In all cases, we account for the effects of Galactic absorption with the multiplicative \texttt{tbabs} model, assuming a constant hydrogen density of $N_\mathrm{H} = 3.51\times 10^{20} \ \mathrm{cm}^{-2}$. For the HETG dataset, we account for potential cross-calibration uncertainties between the HEG and MEG via a multiplicative constant. The last two columns show the goodness-of-fit of the best-fit regimes found when fitting the LETG/HETG spectra, as well as their reduced $C$-stats, defined as $C_\nu = C / \mathrm{DOF}$, where $\mathrm{DOF}$ are the degrees of freedom. We flag the best-fit parameters where the $90\% \ \mathrm{CL}$ given is only approximate with a dagger (${}^{\dagger}$) symbol. The model parameters accompanied by an asterisk (*) underlie parameters which were later seen to be characterised by a bimodal distribution (see Appendix \ref{sec:app_a1_letg_hetg_mcmcRuns} for further details), for which the quoted $90\% \ \mathrm{CL}$ should be considered carefully. Where applicable, we quote the equivalent width (EW) of the \texttt{zgau} model component (quoted at 90\% confidence). The individual LETG/HETG spectra were fitted within their corresponding energy range, given in Tab. \ref{table:bandLumi_and_fluxes}.}
\label{tab:bestfits_relxilllp__from_cstat_min}
\end{table*}

\section{MCMC results on physical parameters from distinct datasets} 
\label{sec:s4_Results_Physical_Params}
A key goal of this study is to constrain the black hole spin parameter ${a}^{*}$ under the assumption that relativistic reflection correctly describes the source spectrum. Individually, the LETG and HETG data indeed provide constraints on the spin but with 90\% confidence ranges that are rather wide. However, it is known that there can be strong covariances between the spin, the disc inclination, and iron abundance, and so the wide range of permitted spins can fundamentally arise from poor constraints on the disc inclination $i$ and/or iron abundance $A_\mathrm{Fe}[Z_\odot]$. With their different energy bands, the LETG and HETG fits have subtle differences in their sensitivity to the inclination and iron abundance. Thus, given that ${a}^{*}$, $i$ and $A_\mathrm{Fe}[Z_\odot]$ should be invariant over human timescales, a joint analysis of the parameter space enabled by reflection on the individual LETG/HETG spectra may permit us to resolve the underlying degeneracies and hence significantly sharpen our spin constraints. Indeed, we find that this is the case (see Fig. \ref{figure:univariate_spin_distributions}). We have verified that the $90\%$ confidence constraints on $a^{*}, i$ and $A_\mathrm{Fe}[Z_\odot]$ we infer as described in the following section are consistent with those inferred by performing a joint LETG/HETG spectral fit (see Appendix \ref{sec:s_b2_3_constraints_fromJointFit} for further details).

In order to produce combined constraints, we firstly performed MCMC analyses of the separate LETG/HETG data, using the covariance matrix of their \texttt{relxill\_lp} best-fits as their initial chain proposal (in Tab. \ref{tab:bestfits_relxilllp__from_cstat_min}), respectively, allowing to more fully explore the parameter space of their reflection model. In both cases, we used the Goodman-Weare algorithm \citep[][]{goodman_weave_mcmc_algorithm_reference_2010} as provided in \textsc{xspec}, with $10^4$ steps and $10^2$ walkers, yielding $10^6$ samples of the posterior. In all cases, we assumed uniformly distributed priors on all free parameter models fitted to the source spectra except for in the case of the ionisation parameter, $\xi$, to which we assigned flat priors in log space. The results of the chains are illustrated by Figs. \ref{figure:LETG_corner_relxilllp+zgau} and \ref{figure:HETG_corner_relxilllp+zgau}, showing the $90\%$ confidence intervals of the relevant free parameters. In fact, several features of the individual LETG/HETG MCMC runs are immediately distinctive, as follows.

Firstly, the distribution of the SMBH spin (${a}^{*}$) and the inclination of the accretion disc ($i$) relative to the rotation axis of the SMBH are in agreement in both MCMC runs. Crucially, both spin distributions are notably skewed towards moderate spin values, i.e. ${a}^{*} \sim 0.4 - 0.6$, while extreme spins are ruled out when adding the upper bound of the $90\%$ credibility ranges resulting from the separate chains, i.e. ${a}^{*}\lesssim 0.86 - 0.92$.

Secondly, the inner disc inclination $i$ is constrained to have moderate values in both chains. Even if the HETG dataset presents stronger features around the iron line when compared to the LETG (see the upper panels in Figs. \ref{figure:hetg_residuals_bestFit_CstatReduction} and \ref{figure:residuals_xspec_letg_bestFit}, respectively), $i$ is more broadly distributed in the HETG chain (see Fig. \ref{figure:HETG_corner_relxilllp+zgau}), with $i \sim {28}^{\circ} - {49}^{\circ}$ at $90\%$ confidence. In this dataset, $i$ is positively correlated or degenerate with the spin parameter ${a}^{*}$, since low inclination values, $\sim {28}^{\circ}$, are coupled to lower values of ${a}^{*}$, and vice-versa.

Thirdly, we note that the distributions for the ionisation state of the inner disc are skewed differently in the LETG/HETG chains (see Appendix \ref{sec:app_a1_letg_hetg_mcmcRuns}). The iron abundance $A_\mathrm{Fe}[Z_\odot]$ is also more broadly distributed in the HETG than in the LETG chain (see Figs. \ref{figure:LETG_corner_relxilllp+zgau} and \ref{figure:HETG_corner_relxilllp+zgau}). Whilst there are no obvious reasons to believe that the metallicity of the accretion disc would readily change between the two observation epochs, it is interesting to note that the lowest bound on $Z$ -- provided by the LETG to be $\sim 0.55Z_\odot$ -- for the accretion disc is \textit{not} consistent with the iron abundance of the circumnuclear material surrounding the immediate vicinity of the AGN inferred by \citet[][]{Reynolds14}, i.e. $\sim0.4Z_\odot$.

Finally, we note that the distributions of \textit{$\Gamma$} and $\mathcal{R}$ are strongly correlated in the LETG chain. This is unsurprising given that enhancing \textit{$\Gamma$} makes the spectrum of coronal emission steeper and naturally induces reflection for an accurate description of the spectrum at higher energies. This does \textit{not} apply to the HETG dataset since it presents distinctive reflection features around the iron band that remain unmodelled by a single power-law continuum, whereas in the LETG spectrum, these features are broader and weaker (see Figs. \ref{figure:hetg_residuals_bestFit_CstatReduction} and \ref{figure:residuals_xspec_letg_bestFit}, respectively). We note that the MCMC chains indicate that \textit{no} significant spectral AGN variability seems to have occurred between the two observation epochs.

\section{Statistical framework to infer merged constraints} 
\label{sec:s5_statisticalFramework_jointConstranits}

We now proceed to describe the merged constraints (i.e. from the distinct LETG/HETG fits) on the invariant properties of the source which ought \textit{not} to change between the LETG and HETG observations.

We identify the set of invariant ``fundamental'' quantities of the system, $\textbf{\textit{f}}$, to be: the dimensionless spin parameter, inclination and iron abundance of the accretion disc, i.e. $\textbf{\textit{f}} = ({a}^{*}, i, A_\mathrm{Fe})$. We identify all other model parameters as ``nuisance'' parameters of the system, $\textbf{\textit{n}}$, in the sense that they can readily change between the different epochs of observation. We label the joint LETG/HETG data by $\{\mathcal{D}_\mathrm{i}\}$, where a given $\mathcal{D}_\mathrm{i}$ (with $i=1, 2$) denotes the individual LETG/HETG datasets.

The MCMC analysis of Sec. \ref{sec:s4_Results_Physical_Params} gives us the posterior for all of the fundamental and nuisance parameters for each of the two observation epochs, $p(\textbf{\textit{f}}, \textbf{\textit{n}}_\mathrm{i}| \mathcal{D_\mathrm{i}})$. For each dataset, one can now marginalise over all nuisance parameters $\textbf{\textit{n}}_\mathrm{i}$ to derive the individual LETG/HETG fit posteriors of the fundamental parameters of the system $p(\textbf{\textit{f}}|\mathcal{D_\mathrm{i}})$ via:\begin{equation}
    \centering 
    \label{equation:posterior_fundamentalParameters_indDatasets}
 p(\textbf{\textit{f}}|\mathcal{D_\mathrm{i}}) \ = \int p(\textbf{\textit{f}}, \textbf{\textit{n}}_\mathrm{i}| \mathcal{D_\mathrm{i}}) \mathrm{d}\textbf{\textit{n}}_\mathrm{i}
\end{equation} where the integral over $\textit{\textbf{n}}_\mathrm{i}$ serves to marginalise over all nuisance parameters for a single dataset. Under the assumption of conditional independence [CI] between the two datasets and assuming uniform priors in $\textbf{\textit{f}}$, the independent fit posteriors $p(\textbf{\textit{f}}|\mathcal{D_\mathrm{i}})$ can be combined via direct multiplication to determine the joint posterior of all fundamental parameters of the system, $p(\textbf{\textit{f}}|\{\mathcal{D_\mathrm{i}}\})$. Mathematically, this corresponds to:
\begin{equation}
\label{equation:posterior_marginalised}
 \centering 
 \mathrm{Merged \ posterior \ [CI] } \  p(\textbf{\textit{f}}|\{\mathcal{D_\mathrm{i}}\}) =  \frac{1}{p(\textbf{\textit{f}})} \prod_{D_\mathrm{i = 1, 2}}  p(\textbf{\textit{f}}|\mathcal{D_\mathrm{i}}),
\end{equation}where $p(\textbf{\textit{f}})$ denotes the prior in $\textbf{\textit{f}}$. We compute Eq. \ref{equation:posterior_marginalised} using the normalisation condition of the merged LETG/HETG posterior over all $\textbf{\textit{f}}$, that is, $\int_{\textbf{\textit{f}}} p(\textbf{\textit{f}}|\{\mathcal{D_\mathrm{i}}\}) \mathrm{d}\textbf{\textit{f}} = 1$.

Operationally, we find the individual LETG/HETG fit posteriors on $\textbf{\textit{f}}$ (Eq. \ref{equation:posterior_fundamentalParameters_indDatasets}) by using \textsc{xspec}'s \texttt{margin} command. For each dataset $\mathcal{D}_\mathrm{i}$, all fundamental parameters were sampled linearly across their full parameter range with $10^2$ steps, i.e. ${a}^{*}\in [-0.99,0.99], \ i\in [{3.5}^{\circ}, 86.5^{\circ}]$, and $A_\mathrm{Fe}\in [0.50,10.0]Z_\odot$.

\begin{figure}
 \center
 \includegraphics[width=0.45\textwidth]{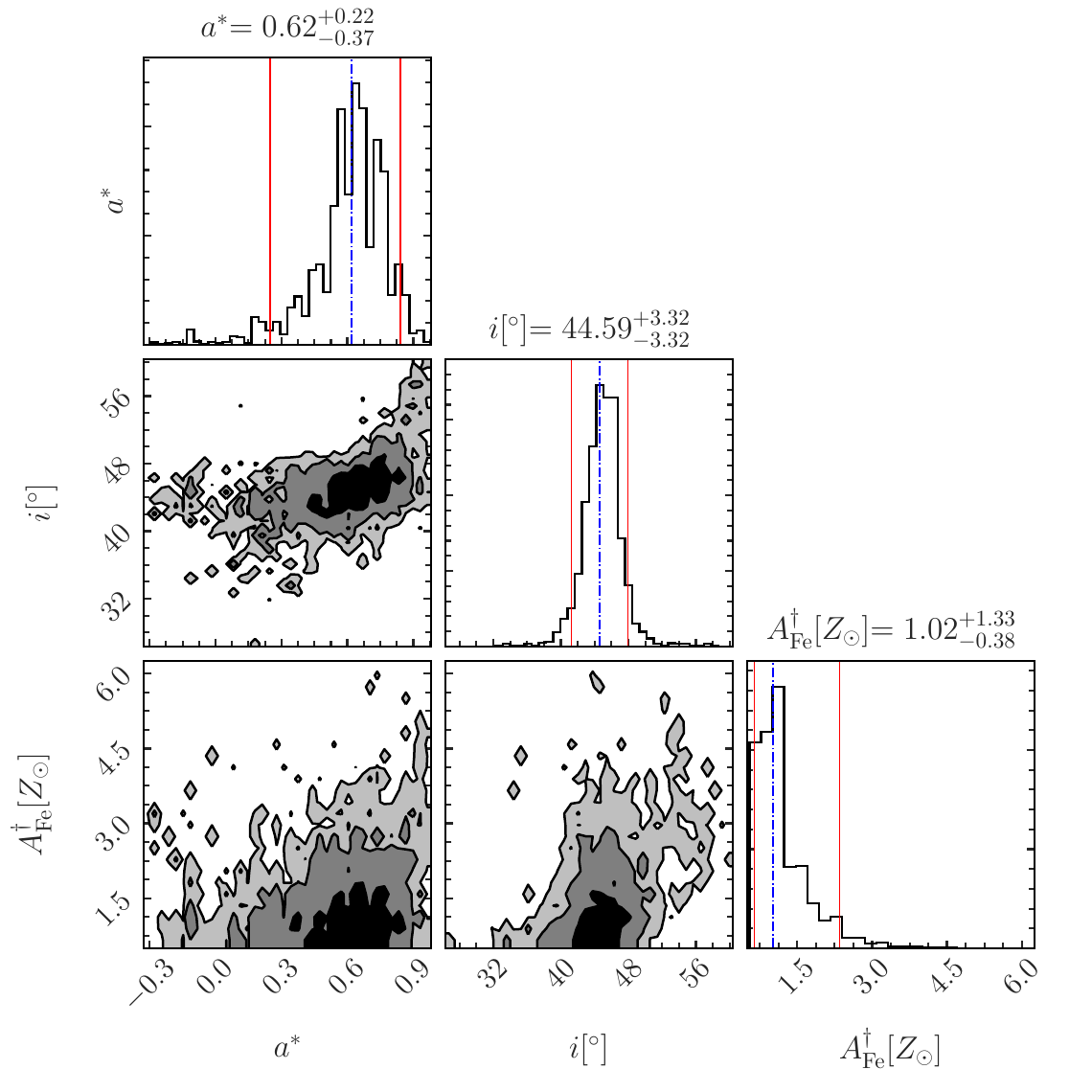}
 \caption{Constraints on system parameters resulting from the merged LETG/HETG dataset. The $90\%$ confidence level ($90\% \ \mathrm{CL}$) drawn from the posterior probability density quoted at the top of each histogram (delimited by solid red lines) is centred at the best-fit median value (dash-dotted blue line) for all fundamental parameters depicted. The parameters flagged with a dagger ($^{\dagger}$) correspond to those whose lower hard limit is comprised within (or close to) their $90\% \ \mathrm{CL}$ lower bound. The contours show the 68\%, 95.5\%, and 99.7\% enclosed probability levels in all covariance plots shown.} 
 \label{figure:mergedDataset_corner}
\end{figure}

The inferred merged constraints on all fundamental parameters of the system $\textbf{\textit{f}}$ found following Equation \ref{equation:posterior_marginalised} is shown by Fig. \ref{figure:mergedDataset_corner}. Indeed, the latter illustrates the apparent sharpening of each fundamental parameter's distribution resulting from having merged the distinct LETG/HETG fit posteriors on $\textbf{\textit{f}}$. Especially apparent is the improvement in the constraints on the spin parameter, with a 90\% credible range of ${a}^{*} \sim 0.25 - 0.84$ from the joint fit posterior. We highlight that this improvement is principally driven by the breaking of the HETG spin-inclination degeneracy with the higher-quality LETG inclination constraint (see Figs. \ref{figure:HETG_corner_relxilllp+zgau} and \ref{figure:LETG_corner_relxilllp+zgau}, respectively).

\section{Discussion}
\label{sec:s6_discussion}
By making the assumption that the X-ray spectral complexities in H1821$+$643 are due to relativistic reflection from the innermost accretion disc, we have shown that the black hole is spinning at a modest rate, ${a}^{*}=0.25 - 0.84$ (90\% credible range). This is a notable improvement on the only previous existing (and tentative) constraint for this object, ${a}^{*}>0.4$ \citep[][]{Reynolds14}. The latter employed \texttt{relxill\_lp} to describe the soft excess on a \textit{Suzaku} spectrum of the source, whereas our study is centred upon statistical evidence of a broadened, fluorescent Fe-$K\alpha$ line that is well described by relativistic reflection of the primary source from the accretion disc.

\begin{figure*}
 \center
 \includegraphics[width=0.75\textwidth]{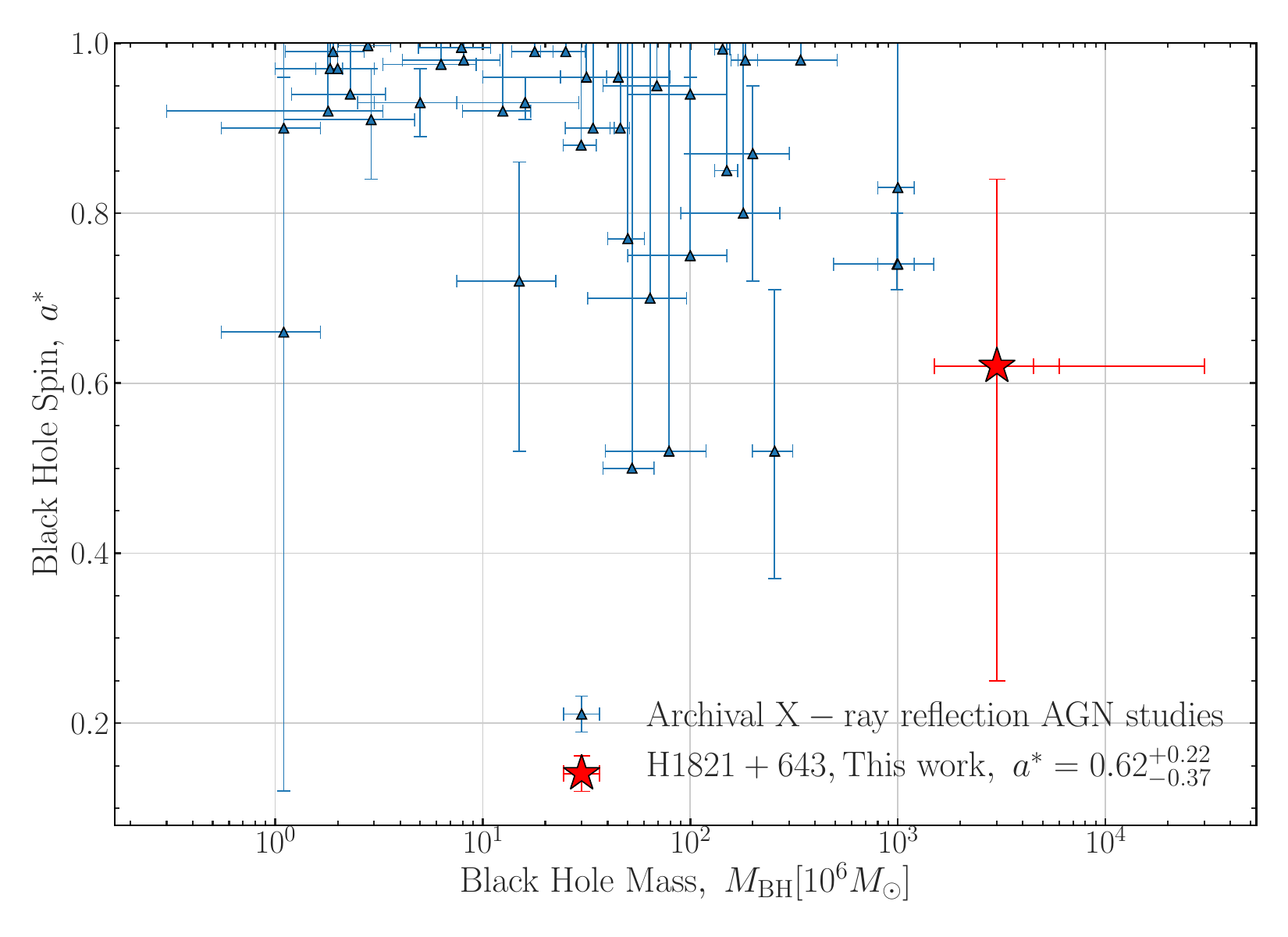}
 \caption{Spin as a function of measured mass for the 32 SMBHs outlined in Table 1 in \citet[][]{reynolds2019ObsBlackHoleSpinsReview}, updated with the inclusion of: I. All but two objects presented in \citet[][]{bambi+21} (refer to discussion in the text); II. The extreme spin constraint for ESO 033-G002 
 of \citet[][]{walton21extreme} -- with its mass estimate; and III. Our result for H1821$+$643 (in red). All spin measurements shown, quoted at $90\%$ confidence, were or have recently been inferred on the basis of X-ray reflection methods. Where available \citep[refer to Tab. 1 in][]{reynolds2019ObsBlackHoleSpinsReview}, all error bars in mass correspond to the $\pm 1\sigma$ level, or are otherwise shown at $\pm 50\%$ significance. Our spin measure for H1821$+$643 is centred at $M_\mathrm{BH} = 3\times 10^9 M_\odot$ \citep[][]{floyd_bcg_mass,Kolman_1991_UV_thinDiscModel,Shapovalova_Monitoring_binBHcandidate,capellupo_1821+643}. Its mass errorbar is overlaid with the upper mass estimate of \citet[][]{Reynolds14}, $6\times 10^9M_\odot$, as well as the mass estimate suggested by \citet[][]{Walker_h1821PressureProfile} on the basis of a Compton cooling flow argument, $3\times {10}^{10}M_\odot$.}
 \label{figure:updated_spin_vs_mass_plot_reynolds20}
\end{figure*}

\begin{figure}
 \center
 \includegraphics[width=0.45\textwidth]{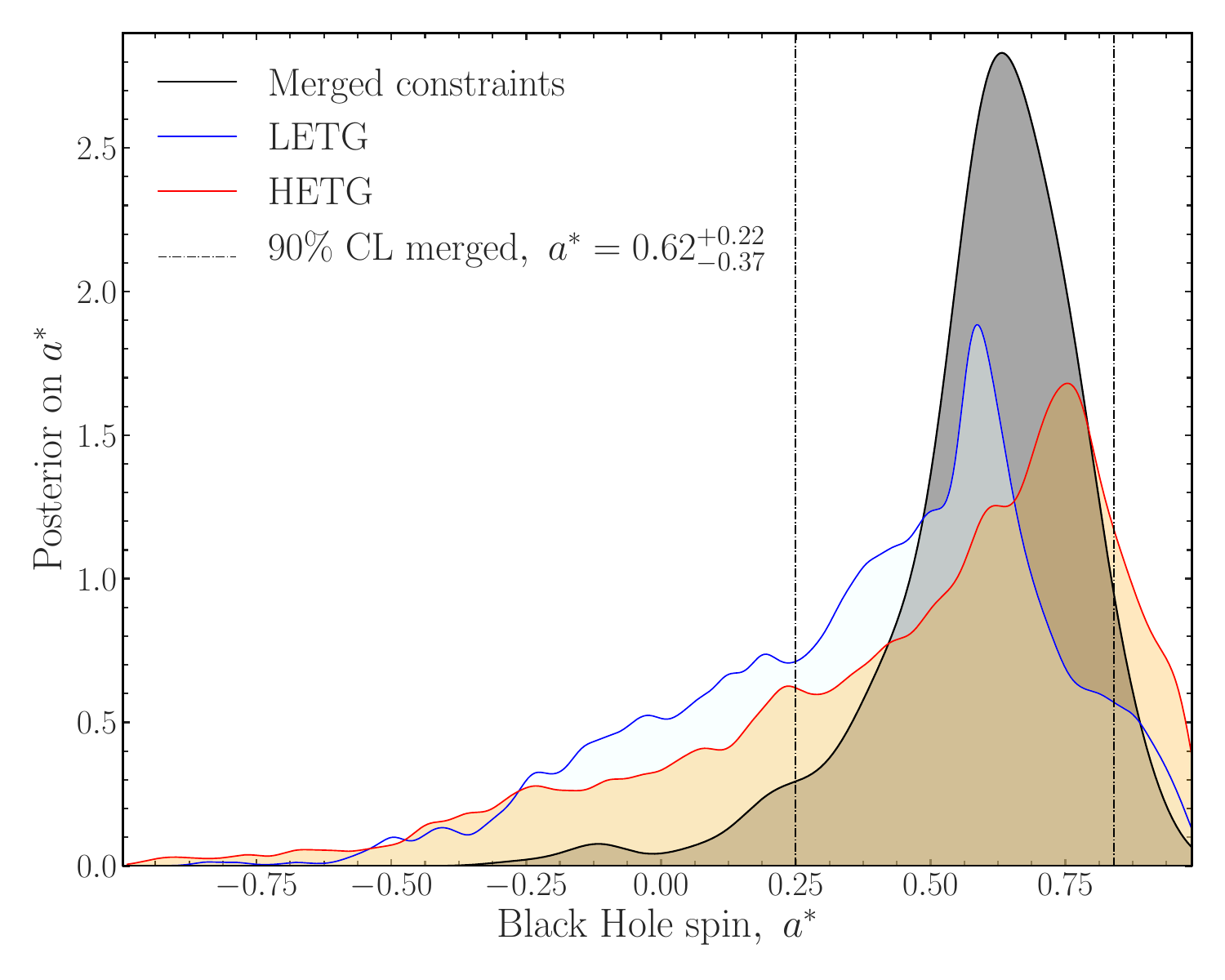}
 \caption{Univariate (coloured) and merged (shaded in grey) black hole spin posteriors based on a gaussian kernel density estimation. The merged distribution was found following the Bayesian procedure described in Sec. \ref{sec:s5_statisticalFramework_jointConstranits}. The dashed-dotted vertical lines delimit the $90\% \ \mathrm{CL}$ on ${a}^{*}$ inferred from the merged distribution.}
 \label{figure:univariate_spin_distributions}
\end{figure}

The constraint we provide on the spin of this object, the most massive SMBH candidate whose angular momentum has been fully constrained (i.e. with a defined upper and lower bound), is perhaps the most interesting result of our work. This is shown by Fig. \ref{figure:updated_spin_vs_mass_plot_reynolds20}, where we include all but two of the spin constraints quoted in \citet[][]{bambi+21}, corresponding to the ultra-fast outflow (UFO) source IRAS 13349+2438 and to the radio-loud quasar 4C 74.26, for the reasons stated below.

Firstly, a combined \textit{Newton-XMM} and \textit{NuSTAR} view of IRAS 13349+243 showed evidence for an UFO in the source spectrum  \citep[see][]{2020_parker_lowSpin, parker_earlier_paper}. UFOs are known to reproduce signatures attributed to relativistic reflection from an ionised accretion disc. Therefore, one may expect the parameter space underlying the description of the iron band in such sources to be notably degenerate. Secondly, we note that the possible broad iron-line radio galaxy 4C 74.26, of an estimated mass $\big ({4.0}^{+7.5}_{-2.5}\big ) \times {10}^{9}M_\odot$ \citep[][]{2002_woo_and_urry_mass_radioloud_4c74.26}, has been tentatively constrained to have a spin of ${a}^{*} > 0.5$, based on the strength of the reflection modelling on a combined \textit{Swift}-\textit{NuSTAR} observation \citep[][]{2017_lohfink_smbh_reflection}. Its timed-average spectrum hinted at the possibility of a recessing inner disc extending up to $\sim$$180R_\mathrm{g}$, which, for instance, one would expect from the disruption of the inner accretion disc by a strong jet event associated with the radio source. In such a scenario, the gravitational imprint of the central SMBH would still be present in the X-ray reflection spectrum from the primary source, although much weaker to that expected for an inner disc fixed at the innermost stable circular orbit.

Our result supports black hole growth scenarios where black holes of masses $\gtrsim 10^8 M_\odot$ would grow following incoherent or chaotic accretion and SMBH-SMBH mergers up to $\gtrsim {10}^{10}M_\odot$. This is predicted by semi-analytic models \citep[][]{king_pringle_chaotic,2014_sesana_semianalytic_SMBHspin_hostgalaxiesproperties, Zhang2019_and_Lu_spin} and by hydrodynamic simulations of cosmic structure formation \citep[][]{bustamante_springel_illustrisTNG}.

 The individual MCMC LETG/HETG analyses suggest that the spectral index of the hard X-ray corona ($\Gamma$) could have marginally changed between the two epochs (see Figs. \ref{figure:LETG_corner_relxilllp+zgau} and \ref{figure:HETG_corner_relxilllp+zgau}). The apparent multi-modal distribution characterising the dimensionless spin parameter of the central SMBH (${a}^{*}$, see Fig. \ref{figure:mergedDataset_corner}) is likely to rise from the marginalisation over degenerate nuisance parameters. We highlight that the inclination of the accretion disc relative to the spin axis of the SMBH is consistent with that predicted in our previous study, $\sim {43}^{\circ}$, where the combined LETG/HETG fit was described with the \texttt{diskline} model \citep[see Sec. 2.2 of][]{sisk21_alps}. 

Merging the separate LETG/HETG chain outputs makes a significant improvement to constraining $A_\mathrm{Fe}$ (see Fig. \ref{figure:mergedDataset_corner} and Figs. \ref{figure:LETG_corner_relxilllp+zgau}, \ref{figure:HETG_corner_relxilllp+zgau}). Indeed, at $90\% \ \mathrm{CL}$, the iron abundance of the accretion disc is allowed to be mildly sub-solar to moderately super-solar, i.e. $Z\sim(0.6 - 2.4)Z_\odot$. We highlight that this lower limit is interestingly discrepant with the iron abundance for the circumnuclear material surrounding the SMBH found in \citet[][]{Reynolds14}, $\sim 0.4Z_\odot$. For this reason, the hypothesis that the cluster-hosted quasar is being fed by an ICM Compton cooling cycle cannot be addressed. We note that our lower bound for $A_\mathrm{Fe}$ in Fig. \ref{figure:mergedDataset_corner} is dominated by the weaker Fe-$K\alpha$ fluorescent line present in the LETG spectrum (see Fig. \ref{figure:residuals_xspec_letg_bestFit}). We also note that a moderate super-solar measurement for iron abundance of the accretion disc could be systematically reduced if the total reflection component were split into: the reflected component attributed to the accretion disc, and that associated with cold, neutral matter surrounding the outer disc. For this reason, in addition to \texttt{relxill\_lp}, we attempted fitting the separate LETG/HETG spectra by using the \texttt{xillver} model \citep[][]{GarciaANDDauser_14_XILLVERnewest} to consider reflection from cold matter surrounding the accretion disc. However, this did \textit{not} improve their goodness-of-fit statistics, given the degeneracies between the two reflection models. When using \texttt{xillver} to model the cold reflection component, we assumed: a torus inclination of $30^{\circ}$ to the normal of the disc; a solar abundance ($\sim Z_\odot$); and a low ionisation state, i.e. $\mathrm{log}(\xi[\mathrm{erg~cm~s^{-1}}]) = 0.2$, for the cold, distant torus surrounding the AGN. We then performed an MCMC analysis of the distinct LETG/HETG datasets following the MCMC setup introduced in Sec. \ref{sec:s4_Results_Physical_Params}. Subsequently, we inferred and merged the independent LETG/HETG fit posteriors on the fundamental parameters of the system, $\textbf{\textit{f}}$, following Sec. \ref{sec:s5_statisticalFramework_jointConstranits}. The $90\%$ confidence bounds on $\textbf{\textit{f}}$ resulting from such an analysis are: ${a}^{*}=0.52^{+0.26}_{-0.45}$, $i[^{\circ}]={43.76}^{+4.98}_{-2.49}$ and $A_\mathrm{Fe}=1.78^{+0.66}_{-0.95}Z_\odot$, and therefore are consistent with those shown in Fig. \ref{figure:mergedDataset_corner}.

On the other hand, we note that previous works \citep[][]{2019_jj_varyingDiscDensity} have shown that varying the density of the accretion disc can significantly affect the iron line profile and measured black hole spin. It is therefore useful to test the sensitivity of our constraints, for instance, to a higher disc density. In particular, we use Fig. 4 of \citet[][]{2019_jj_varyingDiscDensity} to estimate an upper limit on the disc density $n_\mathrm{disc}$ one would expect for H1821$+$643 as a function of $M_\mathrm{BH} \times \dot{m}^{2}$, where $\dot{m}$ is the mass accretion rate in units of the Eddington fraction, $f_\mathrm{Edd}$. Taking $f_\mathrm{Edd} = 0.25$ \citep[][]{Reynolds14} and a black hole mass of $\mathrm{log}(M_\mathrm{BH}/M_\odot) \sim 9$ provides an upper bound of $\mathrm{log}(n_\mathrm{disc}/\mathrm{cm}^{-3}) \lesssim 17$, even if the fraction of the disc energy dissipated into the corona were as high as $\sim 70\%$ (which would be unlikely for such a high-mass black  hole). We then fitted the individual LETG/HETG spectra with the \texttt{const*tbabs*(relxill\_lpD +zgau)} model, where $\texttt{relxill\_lpD}$ describes the reflection spectrum and takes $n_\mathrm{disc}$ as a free parameter, which we set to $10^{17} \ \mathrm{cm}^{-3}$. The best-fitting regimes found for this higher-density model were statistically equivalent to those presented in Tab. \ref{tab:bestfits_relxilllp__from_cstat_min}. The merged $90\%$ confidence bounds on $\textbf{\textit{f}}$ resulting from combining the individual fit LETG/HETG fit posteriors on \textbf{\textit{f}} were found to be consistent with those illustrated by Fig. \ref{figure:mergedDataset_corner}, namely: ${a}^{*}=0.52^{+0.26}_{-0.51}$, $i[^{\circ}]={42.52}^{+3.73}_{-2.08}$ and $A_\mathrm{Fe}=2.16^{+2.76}_{-1.23}Z_\odot$. That is, at $90\%$ credibility, retrograde and maximal spins are excluded and the iron abundance of the inner disc can reach up to $\sim 4.9Z_\odot$.

Additionally, we note that we have restricted our reflection modelling to the consideration of a simple lamppost geometry describing the features present in the LETG/HETG spectra around the iron band (see Figs. \ref{figure:LETG_corner_relxilllp+zgau} and \ref{figure:HETG_corner_relxilllp+zgau}). The spectral models fitted to these datasets also account for: I. A reflected Fe-$6.4 \ \mathrm{keV}$ line from nuclear (cold) matter surrounding the immediate vicinity of the AGN; II. Galactic absorption through the ISM under the assumption of a constant hydrogen column density along the quasar line-of-sight, $N_\mathrm{H} = 3.15 \times {10}^{20} \ \mathrm{cm^{-2}}$; and III. The spectral model fitted to the HETG data also accounts for potential cross-calibration flux uncertainties between the HEG and MEG instruments. We restricted our spectral analysis to energies $\geq1.5 \ \mathrm{keV}$ (observer frame) to avoid complications arising from any soft excess that may be associated with a centrally-concentrated cool ICM.
By assumption, our analysis does \textit{not} include any line-of-sight absorption by an ionised disc wind. Whilst there is no strong evidence for such a component \citep[][]{Oegerle_2000_FUSEObs_confirm_no_outflowEvidence, Fang_2002_WHIMObs_h1821+643,Mathur_2003_WHIM_h1821+643}, these wind signatures could, in principle, imitate several X-ray reflection signatures -- we refer to \citet[][]{parker22_ufo_reflection_degen} and \citet[][]{sim1_outflows,sim2_outflows}. The possible degeneracies on reflection parameters caused by the inclusion of such a disc wind are beyond the scope of this paper.

\section{Conclusions}
\label{sec:s7_conclusions}

In conclusion, we have performed the most rigorous X-ray reflection modelling to date of the powerful radio-quiet cluster-hosted quasar H1821$+$643 using archival \textit{Chandra} grating observations. This has led us to infer the tightest constraints on the spin of the central SMBH, ${a}^{*}=0.62^{+0.22}_{-0.37}$, as well as the inclination $i[^{\circ}]=44.6 \pm 3.3$ and iron abundance $A_\mathrm{Fe}=1.02^{+1.33}_{-0.38}Z_\odot$ of the inner accretion disc (90\% credible ranges). 

The use of \textit{Chandra}’s Low-Energy and High-Energy Transmission Gratings (LETG and HETG, respectively) permit the extraction of a high-resolution intrinsic AGN spectrum free from pile-up where the contamination from cluster emission is minimal. We have employed a total band coverage of $2.0 - 11.0 \ \mathrm{keV}$ (rest frame, $z = 0.299$) to minimise the need for a phenomenological description of the soft excess and to ensure a maximal access to photon counts within the physically relevant iron band. We fit the LETG/HETG spectra separately, given the potential AGN variability between the different times of on-source exposure, finding no significant spectral variability between the two observation epochs. In both cases, we find statistical evidence for reprocessed coronal emission from both the inner accretion disc and colder material in the immediate vicinity of the AGN. We find combined constraints on the fundamental properties of this system by combining the outputs of MCMC chains returning a large sample of possible parameter combinations for both spectra. Our combined constraints are inferred after marginalising over all parameters that can readily change between different observation epochs. 

At $90\%$ confidence ($90\% \ \mathrm{CL}$), we find evidence for a moderate, non-retrograde spin for this massive system, believed to be one of the most massive SMBHs ever detected, i.e. $\sim 3\times [10^9-10^{10}]M_\odot$. We exclude both a maximal, non-rotating and retrograde spin for this object, where ${a}^{*} = 0.25 - 0.84$ at $3\sigma$ significance. This tentatively suggests that accreting systems of similar mass are likely to grow following incoherent accretion and/or SMBH-SMBH merger events. If confirmed, this would importantly distinguish such massive systems from the broader sample of lower-mass AGN, i.e. $\sim (10^6 - 10^7)M_\odot$, whose preferred spins have been characterised to be maximal or near-extreme. 

\section*{Acknowledgements}

We thank the anonymous referee for a constructive and helpful report. We also thank Sam M. Ward for useful discussions on posterior probabilities. J.S-R acknowledges the support from the Science and Technology Facilities Council (STFC) under grant ST/V50659X/1 (project reference 2442592). C.S.R. thanks the STFC for support under the Consolidated Grant ST/S000623/1, as well as the European Research Council (ERC) for support under the European Union’s Horizon 2020 research and innovation programme (grant 834203). J.H.M acknowledges a Herchel Smith Fellowship at Cambridge. R.N.S. acknowledges support from NASA under the Chandra Guest Observer Program (grants G08-19088X and G09-20119X). J.S-R thanks Stefan Heimersheim for relevant advice on the use of \textsc{corner}, and Michael Parker on the use of \textsc{PyXspec}. We would like to thank Dom Walton and Andy Fabian for helpful discussions. We gratefully acknowledge the use of the following software packages: \textsc{astropy} \citep{astropy2013,astropy2018}, \textsc{corner} \citep[][]{corner16_ref}, \textsc{matplotlib} \citep{matplotlib_07} and \textsc{xspec} \citep{XSPEC_1996Arnaud}.

\section*{Data Availability}

The raw X-ray data on which this study is based is available in the public data archives of the Chandra Science Center. The reduced data products used in this work may be shared on reasonable
request to the authors.

\bibliographystyle{mnras}
\bibliography{myPaper}

\appendix

\section{MCMC results from individual LETG/HETG fits}
\begin{figure*}
 \centering
 \includegraphics[width=.95\textwidth]{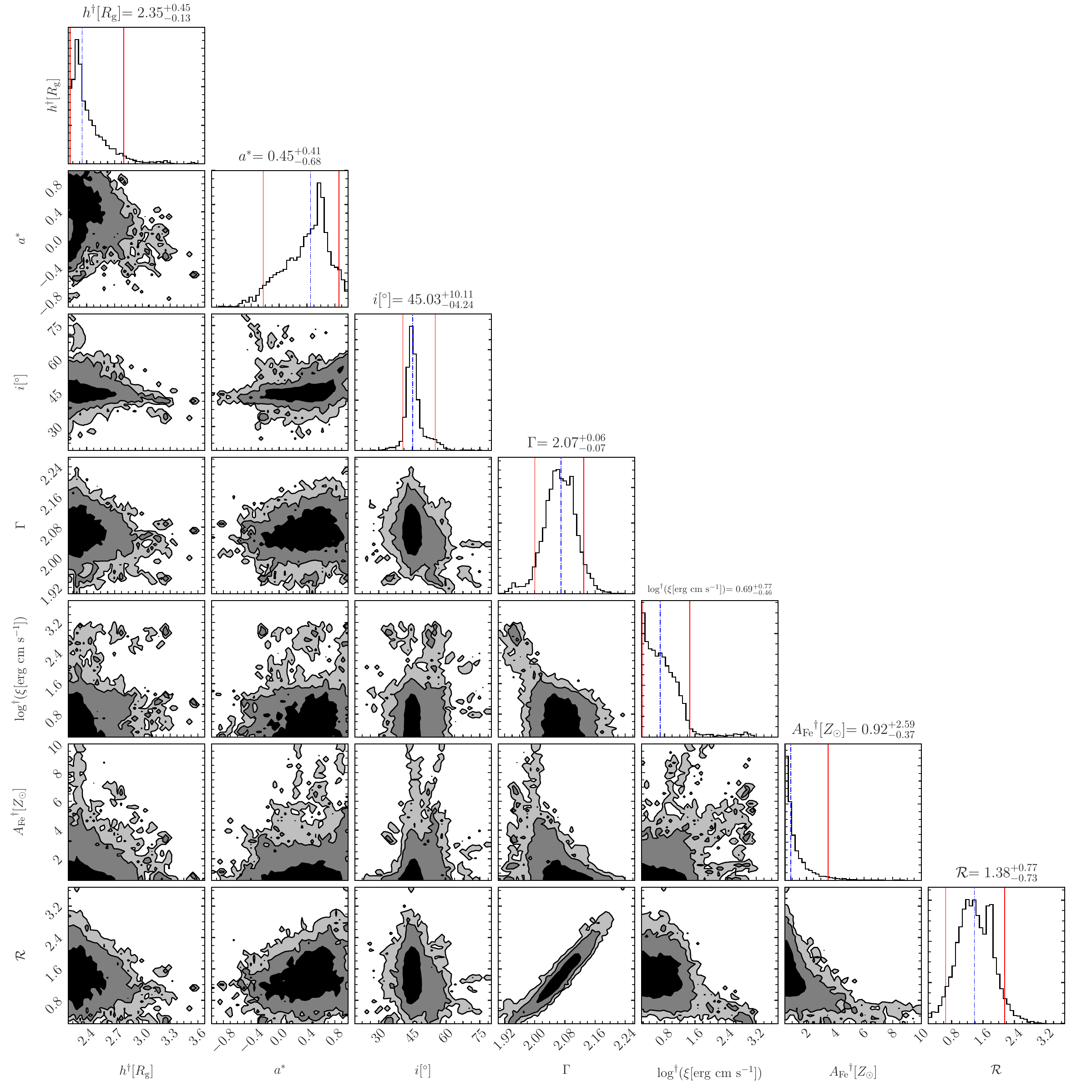}
 \caption{MCMC results inferred when fitting the \texttt{tbabs*(relxill\_lp+zgau)} model to the LETG data, showing the free parameters quantifying the hard X-ray corona incident on the disc ($\textit{$\Gamma$}$) and the relativistic reflection from the disc ($h[R_\mathrm{g}]$, ${a}^{*}$, $i[^{\circ}]$, $\mathrm{log}(\xi[\mathrm{erg~ cm~s^{-1}}])$, $A_\mathrm{Fe}[Z_\odot]$ and $\mathcal{R}$), as listed in Sec. \ref{sec:app_a1_letg_hetg_mcmcRuns}. The $90\%$ confidence level ($90\% \ \mathrm{CL}$) drawn from the posterior probability density quoted at the top of each histogram (delimited by solid red lines) is centred at the best-fit median value (dash-dotted blue line) for all fundamental parameters depicted. The parameters flagged with a dagger ($^{\dagger}$) correspond to those whose lower hard limit is comprised within (or close to) their $90\% \ \mathrm{CL}$ lower bound. The contours show the 68\%, 95.5\%, and 99.7\% enclosed probability levels in all covariance plots shown.}
 \label{figure:LETG_corner_relxilllp+zgau}
\end{figure*}

\begin{figure*}
 \centering
 \includegraphics[width=.95\textwidth]{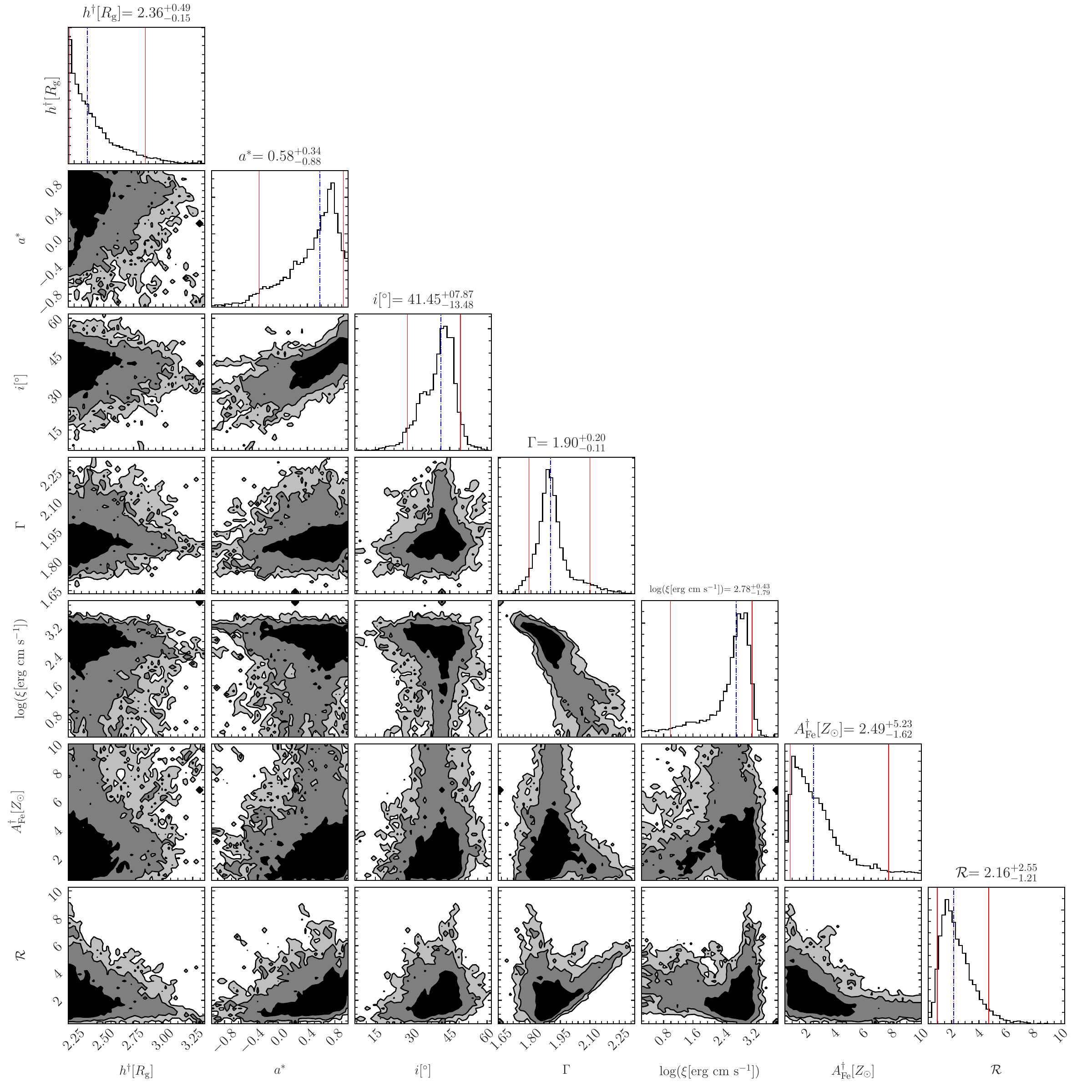}
 \caption{MCMC results inferred when fitting the \texttt{const*tbabs*(relxill\_lp+zgau)} model to the HETG data, showing the free parameters quantifying the hard X-ray corona incident on the disc ($\textit{$\Gamma$}$) and the relativistic reflection from the disc ($h[R_\mathrm{g}]$, ${a}^{*}$, $i[^{\circ}]$, $\mathrm{log}(\xi[\mathrm{erg~ cm~s^{-1}}])$, $A_\mathrm{Fe}[Z_\odot]$ and $\mathcal{R}$), as listed in Sec. \ref{sec:app_a1_letg_hetg_mcmcRuns}. The \texttt{const} model component permits for flux-calibration uncertainties between the HEG/MEG instruments. The $90\%$ confidence level ($90\% \ \mathrm{CL}$) drawn from the posterior probability density quoted at the top of each histogram (delimited by solid red lines) is centred at the best-fit median value (dash-dotted blue line) for all fundamental parameters depicted. The parameters flagged with a dagger ($^{\dagger}$) correspond to those whose lower hard limit is comprised within (or close to) their $90\% \ \mathrm{CL}$ lower bound. The contours show the 68\%, 95.5\%, and 99.7\% enclosed probability levels in all covariance plots shown.}
 \label{figure:HETG_corner_relxilllp+zgau}
\end{figure*}

Figs. \ref{figure:LETG_corner_relxilllp+zgau} and \ref{figure:HETG_corner_relxilllp+zgau} show the results from the MCMC chains performed when fitting the \texttt{tbabs*(relxill\_lp+zgau)} model to the individual LETG/HETG spectra (refer to Secs. \ref{sec:s3.1_fitting_letg} and \ref{sec:s3.2_fitting_hetg}, respectively). In both cases, the neutral hydrogen column density along the quasar line-of-sight was frozen to $3.51 \times {10}^{-20} \ {\mathrm{cm}}^{-2}$ \citep[][]{Reference_forNH}. In the case of the HETG, we accounted for potential flux-calibration corrections between the HEG and MEG. The fitted models consist of the following free parameters: the height $h$[$R_\mathrm{g}$] and reflection fraction $\mathcal{R}$ of the lamppost; the dimensionless spin of the SMBH, ${a}^{*}$; the inclination $i$ and iron abundance $A_\mathrm{Fe}$ of the accretion disc; the ionisation parameter, $\mathrm{log}(\xi[\mathrm{erg~cm~s^{-1}}]$]); and the normalisation of the \texttt{relxill\_lp} and \texttt{zgau} model components (which have \textit{not} been shown in either Fig. \ref{figure:LETG_corner_relxilllp+zgau} or Fig. \ref{figure:HETG_corner_relxilllp+zgau} for simplicity).

We note that the distribution of all values of the normalisation component of the \texttt{zgau} model resulting from the independent LETG/HETG chains are Gaussian and Lorentzian distributed, and predict an average of $(1 - 6)\times 10^{-6}$ and $6 \times 10^{-6} \ \mathrm{photons/cm^{2}/s}$ in the line, respectively. Moreover, the distributions of the normalisation component of the \texttt{relxill\_lp} model component \citep[defined in Appendix A of][]{Dauser_2016_normalisationRelxillDefinition} are positively skewed towards $\sim 2\times {10}^{-3} \ \mathrm{photons/cm^{2}/s}$.

\label{sec:app_a1_letg_hetg_mcmcRuns}

\section{Constraints from a joint LETG/HETG spectral fit}
\label{sec:s_b2_3_constraints_fromJointFit}
To verify our analysis and the results discussed in Sec. \ref{sec:s6_discussion}, we performed a joint LETG/HETG spectral fit using the \texttt{const*tbabs*(relxill\_lp+zgau)} model. All the nuisance parameters of the model associated with the LETG/HETG datasets were fitted freely, whilst the three fundamental parameters of the system, i.e. $\textbf{\textit{f}}=({a}^{*}, {i}^{\circ}, A_\mathrm{Fe}[Z_\odot])$, were tied amongst the two datasets. The best-fit regime was found to be: $C = 3\,991$ for $4\,224$ degrees of freedom (giving a reduced $C$-stat of $C_\nu = 0.945$).

Following Sec. \ref{sec:s4_Results_Physical_Params}, we performed an MCMC analysis of the joint LETG/HETG dataset, assigning a total chain length of $3\,906$ for $256$ independent walkers. Using \textsc{xspec}'s \texttt{margin} command, we then marginalised over all nuisance parameters to find $90\%$ confidence bounds on the three fundamental parameters of the system, $\textbf{\textit{f}}$. These were found to be consistent with the $90\% \ \mathrm{CL}$s on $\textbf{\textit{f}}$ we report in Fig. \ref{figure:mergedDataset_corner}. Crucially, the $90\%$ confidence bound on the black hole spin inferred from the joint fit, ${a}^{*} \sim 0.13 - 0.74$, rules out maximal, retrograde and non-rotating spins and supports our finding that the spin of H1821$+$643 is indeed moderate. 

\label{lastpage}
\end{document}